# Deep learning based supervised semantic segmentation of Electron Cryo-Subtomograms

Chang Liu[1+], Xiangrui Zeng[2+], Ruogu Lin[3], Xiaodan Liang[4], Zachary Freyberg[5], Eric Xing[4], and Min Xu [*2]

[1]Electrical and Computer Engineering Department, Carnegie Mellon University, Pittsburgh, 15213, USA
[2]Computational Biology Department, Carnegie Mellon University, Pittsburgh, 15213, USA
[3]Department of Automation, Tsinghua University, China
[4]Machine Learning Department, Carnegie Mellon University, USA
[5]Departments of Psychiatry and Cell Biology, University of Pittsburgh, USA
[+]Contributed equally

**Abstract**

Cellular Electron Cryo-Tomography (CECT) is a powerful imaging technique for the 3D visualization of cellular structure and organization at submolecular resolution. It enables analyzing the native structures of macromolecular complexes and their spatial organization inside single cells. However, due to the high degree of structural complexity and practical imaging limitations, systematic macromolecular structural recovery inside CECT images remains challenging. Particularly, the recovery of a macromolecule is likely to be biased by its neighbor structures due to the high molecular crowding. To reduce the bias, here we introduce a novel 3D convolutional neural network inspired by Fully Convolutional Network and Encoder-Decoder Architecture for the supervised segmentation of macromolecules of interest in subtomograms. The tests of our models on realistically simulated CECT data demonstrate that our new approach has significantly improved segmentation performance compared to our baseline approach. Also, we demonstrate that the proposed model has generalization ability to segment new structures that do not exist in training data.

## 1 Introduction

Macromolecular complexes are nano-machines that govern molecular processes underlying most aspects of cellular physiology. Elucidating the structures and spatial organization of these complexes is essential for developing a comprehensive understanding of their respective biological processes within the cell. However, such information has been extremely difficult to obtain due to the lack of proper data acquisition techniques. Cellular Electron Cryo-Tomography (CECT) is increasingly proven to be a powerful imaging approach for studying the structures of macromolecular complexes *in situ* [6]. Indeed, recent advances [12, 15, 13, 18] in CECT have enabled 3D visualization of subcellular structures at subnanometer resolution within cells preserved in a near-native state. Such an approach thereby captures the structural and spatial organization information of large macromolecular complexes inside individual cells. However, the systematic computational analysis of macromolecular complexes inside CECT images is very difficult. First, images often have complicated structural contents and a high degree of molecular crowding given the relatively compact nature of the cellular milieu [23]. Second, resolution of these intracellular structures is often limited by a low signal-to-noise ratio (SNR) compared to conventionally frozen purified protein samples, as well as by the missing wedge effect due to limited tilt angle range.

---

[*]Corresponding author email:mxu1@cs.cmu.edu.

Earlier studies that localized macromolecular complexes inside CECT images relied on template matching [4]. However, template matching is limited to only localizing macromolecular complexes of known structures. In order to recover macromolecular structures, reference-free subtomogram[1] averaging and classification [e.g. 6, 3, 17, 21, 9, 22] have been developed. However, accurate similarity measure between subtomograms and subtomogram average is the key to the successful implementation of these approaches. When a subtomogram is extracted from a crowded tomogram, the subtomogram may contain not only the macromolecule of interest, but also neighboring structures. Previously, we have shown that the existence of such neighboring structures could significantly bias the similarity measures, and thereby bias the subtomogram alignment [20], invalidating the structural recovery. Efficient segmentation methods that can accurately detect target regions while filter out neighbor structure regions are needed.

Previously, we developed an unsupervised segmentation approach [20] based on manually designed heuristic rules on over-segmenting, thresholding, and model based clustering. However, such manually designed heuristic rules are far from sufficient in handling the complexity inherent to *in situ* CECT data in the segmentation tasks. For example, the model based clustering step in [20] assumes that the shape of a structural region cluster to be spherical, which does not fit rod or plane like structures. In recent years, deep learning based supervised segmentation has driven advances in semantic segmentation tasks in the computer vision field. In particular, variants of Convolutional Neural Networks (CNN) are applied to semantic segmentation through pixel-wise classification. After large amounts of training data pairs of input image and output segmentation are prepared, a CNN model is able to automatically learn complex segmentation rules hidden inside the training data through a hierarchical combination of convolutional layers. Simple CNN models have been used for supervised 2D segmentation of subcellular structures in CECT [8]. In addition, in our recent work utilizing an unsupervised autoencoder approach for characterizing CECT features [24], we designed an encoder-decoder type CNN model for weakly supervised coarse segmentation. Nevertheless, these simple CNN models have not yet explored the full power of deep learning based methods in 3D CECT semantic segmentation tasks. More accurate and generalizable CNN models therefore need to be designed.

Fully Convolutional Network (FCN) is one of the most popular deep learning based semantic segmentation models. FCN employs convolutional layers, coupled with pooling layers or upsampling layers, to learn coarse information from high layers as well as fine information from low layers. From the combined high and low layers, FCN makes class predictions at every pixel and outputs a segmentation image of the same size as the input image. 2D implementation of FCN, trained end-to-end, pixel-to-pixel, exceeds the current state-of-the-art [14]. In this paper, we proposed a 3D Subtomogram Segmentation Network (SSN3D) inspired by FCN and Encoder-Decoder Architecture [2] for the supervised segmentation of macromolecules of interest in a CECT subtomogram. The basic idea is to merge 3D convolutional layers from pooling and upsampling to combine coarse and fine segmentation information to make predictions about whether the subject 3D voxel belongs to either a background region or to the region of interest (ROI). Besides, the decoder variant of SSN3D has also been explored for learning to map those low resolution to voxel-wise predictions for segmentation.

To demonstrate the utility of our approach, we compared the performance of the proposed method with our previous segmentation approach [20] and evaluated the prediction accuracy on simulated CECT subtomogram data with different image distortion levels. We show that the proposed 3D segmentation network and its variants have significantly improved segmentation performance compared to our previous approach. In addition, we demonstrate that such CNN based segmentation has certain amount of generalization ability: it is possible to use a trained CNN model to segment a structure that does not exist within the training data.

## 2 Method

In this section, we propose 3D segmentation network (SSN3D) and its variants to perform supervised segmentation on subtomograms. Fig.1 shows the inputs and outputs of a 3D segmentation network. Such segmentation can be integrated to a macromolecular structural recovery pipeline that consists of three main steps. First, reference-free particle-picking methods such as the Difference of Gaussian (DoG) approach [16] to detect potential locations of macromolecular complexes. Surrounding regions are then extracted as subtomograms. Second, segmentation is performed on subtomograms to detect ROIs corresponding to the target macromolecules. Background regions, containing noise and fragments of neighboring structures, are masked out. Third, given the segmented subtomograms, subtomogram classification and averaging [e.g. 21] are applied to recover the macromolecular structures.

---

[1]A *subtomogram* is a cubic sub-volume of a tomogram that is likely to contain a single macromolecule.

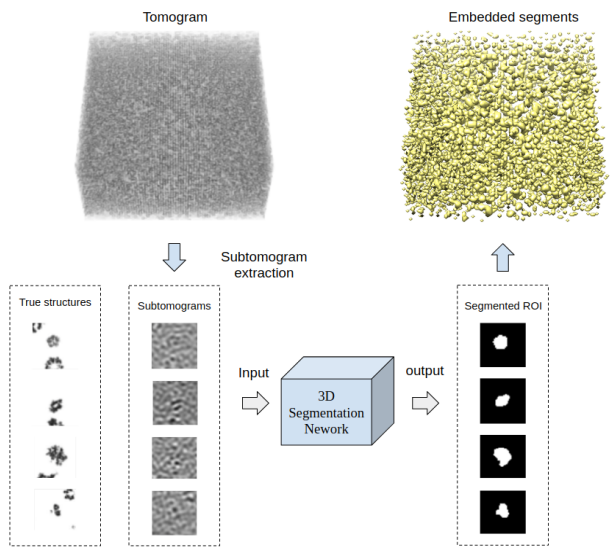

Figure 1: Input and output of SSN3D Network for subtomogram semantic segmentation.

## 2.1 SSN3D Network

Inspired by the Fully convolutional Network adapted from VGG-16[19] classification network, SSN3D is an extension of our recent macromolecule structure classification network, DSRF3D-v2[7] which has achieved good performance for subtomogram classification. For subtomogram segmentation, we have now revised the last fully connected layers into fully convolutional layers to make dense predictions on the voxel level. In addition, SSN3D incorporates skip connection designed from FCN that integrates coarse information from higher feature maps with fine-grained information from lower layers to resolve local ambiguities.

To examine the capability of skip connection in 3D segmentation tasks, two CNN subtomogram segmentation networks are proposed. The first model, named SSN3D-v1 (Fig.2(a)), is a 3D variant of the VGGNet-8 based segmentation network. The input layer of our model accepts 3D subtomograms of size $40 \times 40 \times 40$ voxels. The input layer is connected to three blocks sequentially that consist of two $3 \times 3 \times 3$ 3D convolution layer with stride = 1, and a $2 \times 2 \times 2$ 3D Max pooling layer with stride = 2. Then it is followed by two fully convolutional layers with a dropout rate = 0.7, and three upsampling blocks that consist of two $3 \times 3 \times 3$ 3D convolution layer with stride = 1, and a $2 \times 2 \times 2$ 3D Upsampling layer with stride = 2. All convolutional layers in pooling phase are equipped with a ReLU activation layer while the output layer is equipped with a softmax activation layer. The model does not have skip connection.

The second model, called SSN3D-v2 (Fig.2(b)), has an added skip connection. It differs from the first model only in upsampling phase where the higher layers are combined with the lower pooling layers to integrate the local information. Note that the pooling layers for combination are followed by a $3 \times 3 \times 3$ 3D convolution layer with filter numbers equaling to the number of output classes N.

## 2.2 Encoder-Decoder Extension

In spite of the power and flexibility of the FCN model in 2D datasets, useful global context information is not taken into account by its inherent spatial invariance. Therefore, a 3D encoder-decoder variant of the segmentation model, named SSN3D-ED, is proposed to accurately map low-resolution images to voxel-wise predictions. The architecture of the SSN3D-ED model is shown in Fig.3

Unlike SSN3D-v2 where the number of filters in the decoder part is equal to that of output classes, SSN3D-ED decoder part has the same number of channels as that of encoder. Additionally, we have removed two fully convolutional layers with dropout to decrease the capacity of the model and also added additional convolution layers with ReLU activation before upsampling.

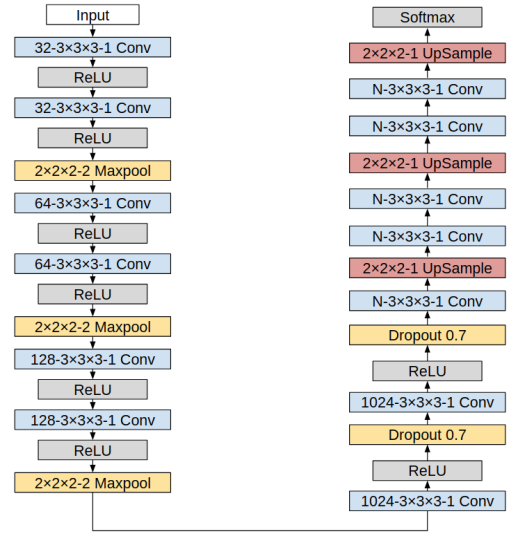

(a) SSN3D-v1 network

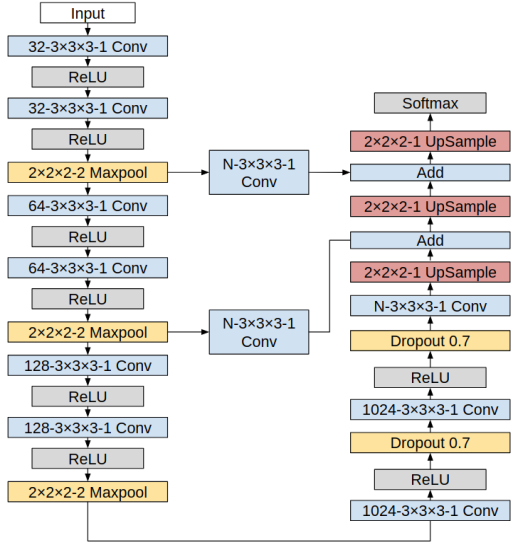

(b) SSN3D-v2 network

Figure 2: Architectures of SSN3D convolutional neural networks.

## 3 Implementation Details

All models were trained and tested on Keras[10] with Tensorflow[1] as the back-end. The experiments were performed on a computer with three Nvidia GTX 1080 GPUs, one Intel Core i7-6800K CPU and 128GB memory.

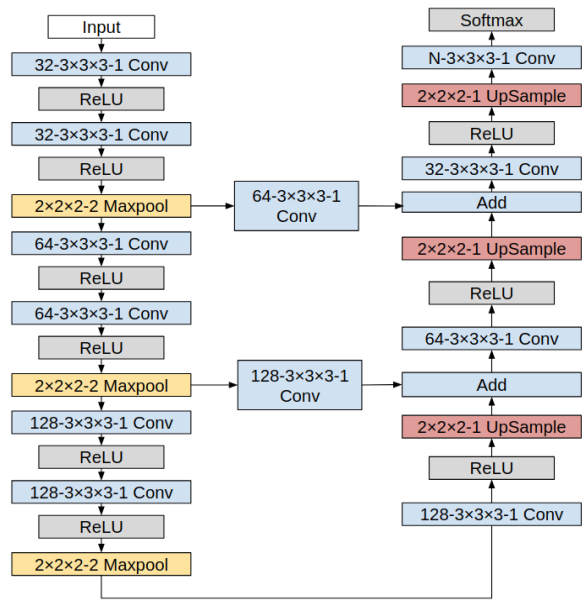

Figure 3: Architectures of EDSSN3D convolutional neural networks.

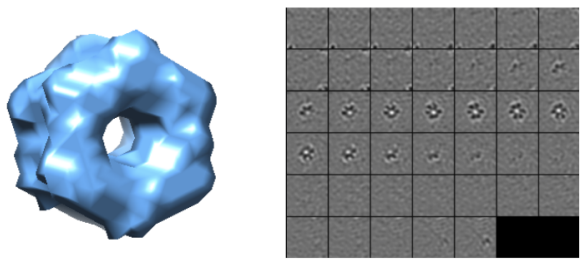

Figure 4: Left: Isosurface of Glutamine synthetase (PDB ID: 2GLS); Right: 2D slices of subtomograms with SNR = 0.5

## 4 Results

### 4.1 Datasets and Evaluation Metrics

**Datasets:** For a convincing assessment of our model, the generation of realistically simulated tomograms is similar to previous work [16] where we used known structures of macromolecular complexes to simulated tomograms by mimicking the actual tomographic image reconstruction process. Fig.4 shows an example of 2D slices of a subtomogram extracted from a simulated tomogram with given SNR. Specifically, we chose 22 distinct macromolecular complexes (Tab.1) from the Protein Databank (PDB) [5] and simulated tomograms of $600 \times 600 \times 300$ voxels and tile angle range $\pm 60°$. Each tomogram that each contains 10000 randomly distributed and oriented macromolecules complexes. Centered on true positions of these macromolecules, the subvolumes of $40^3$ voxels was extracted from each tomogram as input to our model. After removing those outside the boundary of tomograms ($600 \times 600 \times 300$), each dataset contains 3380 simulated subtomograms of 22 structural classes. The SNR of each dataset is estimated using aligned subtomograms containing GroEL complex (PDB ID: 1KP8) according to [11]. Datasets A, B, and C have SNR of 0.5, 0.06, and 0.01 respectively.

**Metrics:** We report two metrics from common semantic segmentation evaluations: pixel accuracy and mean class-wise intersection over union (mIoU).

| PDB ID | Macromolecular Complex |
|--------|------------------------|
| 1A1S | Ornithine carbamoyltransferase |
| 1BXR | Carbamoyl phosphate synthetase |
| 1EQR | Aspartyl-TRNA synthetase |
| 1F1B | E. coli asparate transcarbamoylase P268A |
| 1FNT | Yeast 20S proteasome |
| 1GYT | E. coli Aminopeptidase A |
| 1KPB | PKCI-1-APO |
| 1LB3 | Mouse L chain ferritin |
| 1QO1 | Rotary Motor in ATP Synthase |
| 1VPX | Transaldolase |
| 1VRG | Propionyl-CoA carboxylase |
| 1W6T | Octameric Enolase |
| 1YG6 | ClpP |
| 2AWB | Bacterial ribosome |
| 2BO9 | Human carboxypeptidase A4 |
| 2BYU | M.tuberculosis Acr1(Hsp 16.3) |
| 2GHO | Thermus aquaticus RNA polymerase |
| 2GLS | Glutamine Synthetase |
| 2H12 | Acetobacter aceti citrate synthase |
| 2IDB | 3-octaprenyl-4-hydroxybenzoate decarboxylase |
| 2REC | RECA hexamer |
| 3DY4 | Yeast 20S proteasome |

Table 1: The experimental macromolecular complexes used for tomogram simulation and semantic segmentation.

## 4.2 Segmentation performance

**Optimization:** We randomly split the dataset into 3042 subtomograms for training, and 338 subtomograms for testing. 10 % of training dataset were used for validation. All CNN models were trained using Adam with the learning rate of $10^{-3}$ under a decay factor of $1e^{-8}$ and exponential decay rates $\beta 1 = 0.9$ and $\beta 2 = 0.99$ to minimize the categorical cross-entropy cost function. Adam training was performed with a batch size of 128 for 200 epochs, while the early stopping criterion was set such that the training would stop and be saved if the validation dataset did not improve for 10 consecutive epochs in terms of pixel accuracy.

**Evaluation:** Using our datasets, we compare the segmentation performance of three proposed CNN models including SSN3D-v1, SSN3D-v2 and SSN3D-ED, with our previously designed method [20] as baseline. The performance results are summarized in Table 2. It can be seen that all CNN models achieved significantly higher pixel-wise accuracy and mIoU than our baseline method.

For the illustration of the capability of skip connection in 3D voxel level segmentation, SSN3D-v2 has a slightly higher pixel accuracy (99.30%) and an apparent improvement on mean IoU (85.30%) compared to SSN3D-v1. From the comparison, we can infer that combining coarse information from higher layers and fine-grained information from lower layers remarkably enhances the segmentation performance of CNN model in 3D voxel level dataset.

To verify the hypothesis that decoder variants also work well in 3D case, SSN3D-ED is proposed. From the Tab.2, SSN3D-ED achieves similar but slightly higher pixel accuracy and mean IoU compared to SSN3D-v2. More importantly, its model capacity is just the half that of SSN3D-v2. Therefore, we can conclude that the decoder variant also helps 3D voxel level segmentation. Apart from this, we have also tried various deeper encoder architectures under the similar decoder design, but none of them outperform SSN3D-ED. Thus, encoder-decoder architecture in 3D subtomogram segmentation still needs to be further explored.

| Method   | Pixel Acc. | mIoU  | Capacity  | SNR  |
|----------|------------|-------|-----------|------|
| Xu[20]   | 91.87      | 27.40 |           | 0.5  |
| SSN3D-v1 | 98.28      | 70.10 | 2,042,158 | 0.5  |
| SSN3D-v2 | 99.30      | 85.30 | 2,042,342 | 0.5  |
| SSN3D-ED | **99.35**  | **86.40** | 895,746 | 0.5 |
| SSN3D-ED | 98.84      | 83.04 | -         | 0.06 |
| SSN3D-ED | 97.94      | 73.78 | -         | 0.01 |

Table 2: Segmentation performance on simulated subtomograms.

In order to test the robustness of our model, which achieves good segmentation performance even for relatively low SNR tomograms, we trained our models on datasets of different SNRs. The results are shown in Table 2. It is proved that our best model achieves no less than 97 % accuracy and 73 % mIoU even under the lowest SNR setting.

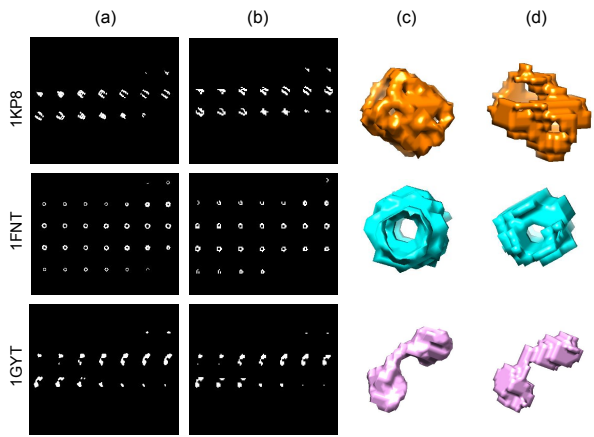

Figure 5: Examples of ground truth and prediction pairs: (a) 2D slices of ground truth (b) 2D slices of predicted segmentation (c) 3D isosurfaces of ground truth (d) 3D isosurfaces of predicted segmentation.

### 4.3 Generalization ability analysis

To illustrate that our best proposed CNN model has certain generalization ability, we split our dataset A into training and testing sets based on the leave-one-out rule. Specifically, 1 out of 22 macromolecular complexes classes is used as a test structure, and we trained our SSN3D-ED model on the dataset that only contains the remaining 21 structural classes. The segmentation performance of all 22 structures are summarized in Table 3. The segmentation of each new structure achieved pixel accuracy of at least 96%, and mIoU of at least 72%, showing good generalization ability.

## 5 Conclusion

In this paper, a novel 3D CNN Segmentation network (SSN3D) and its variants are presented to significantly improve deep learning based subtomogram segmentation compared to our previous method [20]. SSN3D-ED is our best model among those proposed methods, achieving 99.35 % pixel accuracy and 86.40 % mean IOU. SSN3D-ED demonstrated its ability to segment subtomograms with realistically simulated low SNR subtomograms that may contain neighbor structures due to molecular crowding. In addition, we demonstrate that such CNN based segmentation is supervised, and has certain generalization ability: it is possible to use a trained CNN model to segment a structure that does not exist in the training data. For future work, the proposed CNN architectures remain to be further optimized for improving segmentation performance on tomograms with lower SNR.

| PDB ID | Pixel Acc. | mIoU |
|---|---|---|
| 2AWB | 96.57 | 81.47 |
| 1FNT | 98.11 | 87.59 |
| 1GYT | 98.51 | 86.95 |
| 1BXR | 97.45 | 77.02 |
| 1QO1 | 98.45 | 75.23 |
| 1KP8 | 98.38 | 88.85 |
| 3DY4 | 98.65 | 88.08 |
| 1VPX | 98.99 | 88.15 |
| 1EQR | 99.20 | 83.29 |
| 2GLS | 98.83 | 88.12 |
| 2GHO | 99.22 | 86.45 |
| 1VRG | 99.23 | 86.89 |
| 2H12 | 99.26 | 86.45 |
| 1YG6 | 99.22 | 85.83 |
| 2REC | 98.97 | 72.43 |
| 2BO9 | 99.60 | 86.16 |
| 2BYU | 99.44 | 86.72 |
| 2IDB | 99.43 | 85.07 |
| 1F1B | 99.46 | 81.71 |
| 1W6T | 99.67 | 85.46 |
| 1A1S | 99.74 | 80.66 |
| 1LB3 | 99.78 | 79.61 |

Table 3: Generalization performance of our best proposed model SSN3D-ED (SNR = 0.5)

**Acknowledgements**: We thank Dr. Robert F. Murphy and Mr. Bo Zhou for suggestions. This work was supported in part by U.S. National Institutes of Health (NIH) grant P41 GM103712. M.X acknowledges support of the Samuel and Emma Winters Foundation.